\documentclass{article}
\usepackage{ijcai26}

\usepackage{times}
\usepackage{soul}
\usepackage{url}
\usepackage[utf8]{inputenc}
\usepackage[small]{caption}
\usepackage{xcolor}
\usepackage{graphicx}
\usepackage{amsmath}
\usepackage{amsthm}
\usepackage{booktabs}
\usepackage{algorithm}
\usepackage{algorithmic}
\usepackage[switch]{lineno}
\usepackage{latexsym}
\usepackage[hidelinks]{hyperref}
\usepackage{cleveref}
\usepackage{listings}

\usepackage{tikz}
\usetikzlibrary{arrows.meta,shapes.geometric,decorations.pathreplacing,positioning,calc}
\usepackage{pgfplots}
\usepackage{placeins}
\newtheorem{example}{Example}

\title{Agentic SPARQL: Evaluating SPARQL-MCP-powered \\ Intelligent Agents on the Federated KGQA Benchmark}

\author{
Daniel Dobriy$^1$\thanks{Corresponding author. Email: daniel.dobriy@wu.ac.at}
\and
Frederik Bauer$^1$
\and
Amr Azzam$^1$
\and
Debayan Banerjee$^2$
\and
Axel Polleres$^{1,3}$\\
\affiliations
$^1$WU Vienna, Vienna, Austria\\
$^2$Leuphana Universit\"at L\"uneburg, L\"uneburg, Germany\\
$^3$Complexity Science Hub Vienna, Austria
}

\begin{document}

\maketitle

\begin{abstract}
    Standard protocols such as the Model Context Protocol (MCP) that allow LLMs to connect to tools have recently boosted ``agentic'' AI applications, which, powered by LLMs' planning capabilities, promise to solve complex tasks with the access of external tools and data sources. In this context, publicly available SPARQL endpoints offer a natural connection to combine various data sources through MCP by (a) implementing a standardised protocol and query language, (b) standardised metadata formats, and (c) the native capability to federate queries. In the present paper, we explore the potential of SPARQL-MCP-based intelligent agents to facilitate federated SPARQL querying: firstly, we discuss how to extend an existing Knowledge Graph Question Answering benchmark towards agentic federated Knowledge Graph Question Answering (FKGQA); secondly, we implement and evaluate the ability of integrating SPARQL federation with LLM agents via MCP (incl.\ endpoint discovery/source selection, schema exploration, and query formulation), comparing different architectural options against the extended benchmark. 
    Our work complements and extends prior work on automated SPARQL query federation towards fruitful combinations with agentic AI.
\end{abstract}

\section{Introduction}\label{sec:intro}

In recent years, Large Language Models (LLMs) have impacted nearly every industry and many aspects of daily life, prompting transformations in the Web's ecosystem (such as blurring the boundaries between `search' and `question answering') 
as well as motivating research focused on addressing their major limitations: hallucinations, bias, knowledge cut-offs, reasoning deficiencies and lacking explainability/interpretability \cite{matarazzo2025survey}. A particular challenge, but also opportunity, in this context lies in the combination of LLMs with external, structured data sources, which has been addressed in different, complementary research strands:

\textbf{RAG \& Text-to-Query.} Firstly, Retrieval-Augmented Generation (RAG), a technique of retrieving and injecting relevant and structured external knowledge into an LLM's model context~\cite{lewis2020retrieval}, has been proposed and widely adopted in various forms.
While RAG is often used in combination with textual embeddings and vector databases to map and lookup in relational or graph data most related items to a prompt by simple ``top-k lookups'' from a database in terms of vector similarity \cite{karpukhin2020dense}, another common approach to retrieval is text-to-query that (i) translates LLM prompts or contexts to database queries, with, most prominently, Text-to-SQL \cite{liu2025survey} or Text-to-SPARQL \cite{10.1145/3477495.3531841,dabramo-etal-2025-investigating} as central components of Knowledge-Graph-Question-Answering (KGQA) systems, and (ii) feeds the query results, directly or via templated result-to-text translations, back to the LLM. 

\textbf{Agentic AI.} More recently and significantly boosting the base idea of mostly hard-coded RAG pipelines, the Model Context Protocol (MCP) has been proposed as an open standard for context management in ``agentic'' settings.\footnote{See \url{https://www.anthropic.com/news/model-context-protocol}} At its core \cite{anthropic2025specification}, besides server-provided resources (contextual data), prompts (templated workflows), and tools (executable methods), the main novelty over hard-coded RAG pipelines lies in that the standard agent protocol and standard reasoning/planning capabilities within modern LLMs, such as Reason+Act (ReACT) \cite{yao2023react}.

\textbf{Federated Querying.} Use cases on agentic AI in the context of querying databases explored so far in the literature have mostly focused on interactions with \emph{single} (e.g. relational) databases and endpoints: while tasks such as schema exploration, and query formulation over single databases have been successfully delegated to such agents, the power of exploiting \emph{federated} query endpoints in agentic AI applications has received less attention so far. Yet, in terms of APIs potentially usable for agentic AI pipelines, W3C standard recommendations such as the Resource Description Framework (RDF) \cite{rdf11-concepts} along with the SPARQL Protocol and RDF Query Language (SPARQL) \cite{harris2013sparql} already provide standardized API access to federated graph data, as primary standards for publishing and querying linked open data (LOD) and FAIR data \cite{wilkinson2016fair}. Additionally, SPARQL 1.1 introduced support for \emph{federated querying} natively in the standard, through the SERVICE operator~\cite{harris2013sparql}, as well as dedicated RDF vocabularies to describe and advertise SPARQL endpoints~\cite{williams2013sparqlservicedescription,bohm2011creating}. The availability of such standards has motivated research on various aspects of federated SPARQL query processing, addressing challenges such as endpoint discovery, query decomposition, and query planning \cite{fedx,DBLP:conf/semweb/AcostaVLCR11}, including the definition of benchmarks such as FedBench~\cite{schmidt2011fedbench}.

\textbf{Towards ``Agentic SPARQL''.} As such, flexibly integrating SPARQL endpoints and query federation into agentic LLM applications seems a natural next step towards allowing agents to integrate federated graph data served via SPARQL endpoints. In the present paper, we will refer to \emph{Agentic SPARQL} as the combination of agentic AI, and federated SPARQL querying capabilities providing agents with the access to the Web of Data. In this context, we see the following main open challenges, which we aim to address in our work:

\begin{description}
\item[C1: Interface Heterogeneity.] ``Endpoints'' may range from lower level (below-SPARQL) interfaces like RDF dumps only, to triple-pattern based interfaces and, finally, to "full" SPARQL endpoints. Azzam et al.\cite{azza-etal-2024SWJ} have recently characterized and reviewed these different interfaces in a uniform formalization.

\item[C2: Uneven SPARQL 1.1 Support.] Closely related to C1, even endpoints that natively support the standard SPARQL protocol may vary greatly in terms of uneven SPARQL 1.1 support, in terms of language features (e.g. aggregates, supporting VALUES or SERVICE clauses), or limiting the result sizes, all of which require bespoke strategies for federation.


\item [C3: Metadata Heterogeneity.] Catalogues of source metadata to inform endpoint discovery, schema exploration and query planning are largely absent and neither fully standardized, and as a consequence efficient source selection remains difficult, often leading to slow query executions due to overestimation of relevant sources~\cite{saleem2016fine}.

\item [C4: Latency, Availability and Timeouts.] SPARQL endpoints are sometimes blocked or exhibit high latency, timeouts, and even up non-availability due to high concurrent server loads or complexity of queries, which requires federation engines to support non-blocking joins~\cite{saleem2016fine}. In general, the variable availability of endpoints and non-predictability of timeouts largely affects query planning (and potentially also schema exploration).
\end{description}

Overall, C4 may affect the effectiveness of predominant ``plan-and-execute'' approaches, under circumstance where latency and up-to-date information changes dynamically during execution, or query (parts) are not executable in reasonable time.

\begin{description}
\item[C5: Query Formulation.] Orthogonal to the above challenges, when starting from a (natural language) question, query formulation accuracy, is still not entirely solved. Firstly, this depends on the ability of (small) LLMs to produce \emph{syntactically} correct queries and, secondly, on the semantic challenges in ``knowing'' the correct identifiers within queried KGs.
\end{description}

Thus, while impressive progress in regards of query formulation for specific SPARQL endpoints such as Wikidata~\cite{liu2024spinachsparqlbasedinformationnavigation} have been made, the challenge of getting from natural language queries to results is aggravated in dynamic federated environments where neither the relevant endpoint URLs nor the schema is known upfront, as illustrated in \Cref{ex:1}:

\begin{example}\label{ex:1}
As an illustrating example, prompted to \textsf{gpt-oss-20b}\footnote{See \url{https://openai.com/index/introducing-gpt-oss/}}, let us attempt to generate a federated SPARQL query: 

\begin{quote}
    ``Write a federated SPARQL query that retrieves all of Tim Berners-Lee's publications from \underline{DBLP}, whose DOIs also appear on \underline{Wikidata}.''
\end{quote}

\begingroup
\normalfont
\begin{lstlisting}[basicstyle=\scriptsize\ttfamily,escapeinside=||]
PREFIX dblp:   <|\colorbox{magenta!25}{http:}|//dblp.org/rdf/schema#>
PREFIX bibo:   <http://purl.org/ontology/bibo/>
PREFIX wdt:    <http://www.wikidata.org/prop/direct/>
SELECT DISTINCT ?dblpPub ?title ?doi ?wikidataItem
WHERE {
 SERVICE <|\colorbox{magenta!25}{https://dblp.org/sparql}|> {
   ?dblpPub a bibo:Article ;
           dblp:|\colorbox{magenta!25}{author}| ?author ;
           dblp:title  ?title ;
           bibo:|\colorbox{magenta!25}{doi}| ?doi .
           ?author dblp:|\colorbox{magenta!25}{name}| "Tim Berners-Lee".}
SERVICE <https://query.wikidata.org/sparql> {
   ?wikidataItem wdt:P356 ?doi . } }
\end{lstlisting}
\endgroup
\end{example}

As the example shows, both the endpoint address as well as the schema (incl. the prefix) were predicted erroneously. Furthermore, while on the surface level the query looks reasonable, a naive query planner, executing this query on a local engine capable of running \texttt{SERVICE} queries, could likely find a better plan than retrieving and locally joining all DOIs (320,960) from Wikidata and locally joining them with the (far fewer) DOIs retrievable for Tim Berners-Lee from DBLP, or resp. we need to rely on the query optimization capabilities of an overall local engine that executes the resulting query. 

Our main hypothesis is that a combined agentic approach can help address these challenges, but this combination needs to go beyond current works: while existing approaches (see \Cref{sec:related}) have already introduced SPARQL querying capabilities into MCP, these have thus far been restricted to single endpoints and have not addressed dynamic endpoint discovery, schema exploration or federated querying. Also, KGQA benchmarks for SPARQL query formulation do so far not cover federation; consequently, we will need to combine existing KGQA and federation benchmarks. In our work, we aim to take a first crucial step by both (i) establishing such a combined benchmark and (ii) examining how to enable ``agentic SPARQL" with an implementation, based on MCP, which we (iii) evaluate using different state-of-the-art LLMs.


The remainder of this work is structured as follows: \Cref{sec:preliminaries} introduces preliminary concepts; \Cref{sec:sparql-mcp} presents our SPARQL-MCP server; \Cref{sec:fkgqa-benchmark} describes the combined federated KGQA benchmark; \Cref{sec:evaluation} and \Cref{sec:results} present the evaluation setup as well as results; \Cref{sec:discussion} and \Cref{sec:related} discuss limitations and related works; finally, \Cref{sec:conclusions} concludes the paper and gives an overview of future work.

\section{Preliminaries}\label{sec:preliminaries}

\begin{figure*}[h]
    \centering
    \resizebox{\textwidth}{!}{%
    \begin{tikzpicture}[
        x=1cm,y=1cm,
        every node/.style={font=\scriptsize},
        box/.style={draw, rounded corners, align=center, inner sep=4pt, minimum width=3.8cm, minimum height=0.9cm},
        rectbox/.style={draw, align=center, inner sep=4pt, minimum width=3.8cm, minimum height=0.9cm},
        clientbox/.style={draw, rounded corners, line width=0.3pt, align=center, inner sep=4pt, minimum width=3.8cm, minimum height=0.9cm},
        db/.style={cylinder, draw, shape border rotate=90, aspect=0.12, align=center, inner sep=4pt, minimum width=3.8cm, minimum height=0.9cm}]
        \tikzset{>=latex,
            mcp/.style={<->, very thick, dashed},
            sparql/.style={<->},
            subq/.style={<->, dashed}}

        \node[clientbox] (client) at (0,0) {Client\\(SPARQL-MCP caller)};
        \node[box] (mcp)    at (5,0) {SPARQL-MCP server\\(tools: schema, federated query)};
        \node[db]  (a)      at (10,0) {Remote SPARQL\\endpoint A};
        \node[rectbox] (fed)    at (5,-1.5) {SPARQL federation endpoint};
        \node[db]  (b)      at (10,-1.5) {Remote SPARQL\\endpoint B};
        \node[db]  (catalogue) at (0,-1.55) {Internal catalogue\\(endpoint registry)};

        \draw[mcp] (client) -- (mcp);

        \draw[sparql] (mcp) -- (a);
        \draw[sparql] (mcp) -- (b);

        \draw[sparql] (mcp) -- (fed);
        \draw[subq] (fed) -- (a);
        \draw[subq] (fed) -- (b);

        \draw[] (mcp) -- (catalogue);

        \begin{scope}[yshift=-2.5cm]
            \draw[mcp] (-0.9,0) -- (1.1,0);
            \node[right] at (1.3,0) {\textbf{MCP}};
            \draw[sparql] (2.6,0) -- (4.4,0);
            \node[right] at (4.5,0) {SPARQL (w. SERVICE)};
            \draw[subq] (8.1,0) -- (9.9,0);
            \node[right] at (10,0) {Subqueries};
        \end{scope}

    \end{tikzpicture}}
    \caption{Architecture: the client communicates with the SPARQL-MCP server via the MCP protocol. The server includes the internal catalogue (endpoint registry) as a component, and issues SPARQL queries either directly to remote endpoints (A, B) or to a federation endpoint which decomposes and forwards subqueries to A and B for multi-SERVICE queries.}
    \label{fig:architecture}
\end{figure*}
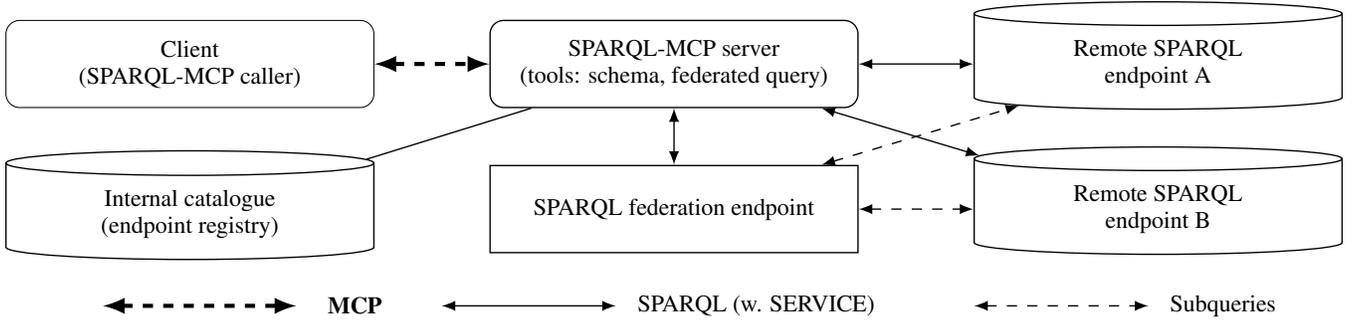

In the following, we assume familiarity with RDF, the SPARQL query language and protocol, and the MCP request/response protocol, and will only recapitulate briefly the parts relevant for our paper.

\textbf{RDF \& SPARQL} provide standards to publish and query Web-accessible datasets consisting of (RDF) triples, that could be viewed as subject-predicate-object statement, or, likewise, as edges of a directed, labelled graph:
let $G$ be the RDF graph of a dataset, i.e., a set of RDF triples $(s_d,p_d,o_d) \in (I\cup B)\times I\times (I\cup B \cup L)$ with $I$, $B$,$L$ denoting IRIs, blank nodes, or literals, resp. A \emph{triple pattern} is an RDF triple $(s_q,p_q,o_q)$ where also variables from set $V$ are allowed in either position. A \emph{basic graph pattern} (BGP) is a finite set of such triple patterns, and can be read as a conjunctive query. As illustrated in \Cref{ex:1}, BGPs can be written in Turtle  syntax~\cite{rdf11-concepts} in SPARQL, and be composed to more complex patterns, incl. a SERVICE $e$ $P$ operator that allows to delegate the execution of a SPARQL query pattern $P$ to an \emph{endpoint} $e\in \mathcal{E}$ (the set of all endpoint URIs. A UNION($P_1,\ldots,P_n$) operator can be used express unions of conjunctive queries. For details of other operators such 
aggregating subqueries (e.g. COUNT, SUM, etc.), as well as FILTER expressions and LIMIT/OFFSET operators we refer to \cite{buil-etal-2013-fedsparql-JWS}. 


We herein treat "endpoint" as an HTTP/HTTPS interface which accepts and answers SPARQL queries using the standard SPARQL protocol (and assume lower-level interfaces such as triple pattern fragments as simply endpoints only accepting restricted SPARQL queries, cf. \cite{azza-etal-2024SWJ}). Conveniently, the SPARQL protocol offers a JSON result format amenable to MCP: we assume all endpoints in $\mathcal{E}$ to support such JSON results. 

For the sake of our paper and benchmark, we will assume that any natural language question $q_n$ we start from can be translated (minimally) to a single BGP query, and \emph{query federation} involves splitting up this pattern to either (i) disjoint SERVICE sub-patterns delegated to different endpoints (\emph{conjunctive split}), or (ii) creating UNIONs-concatenated copies of SERVICE-annotated patterns delegated to different endpoints (\emph{disjunctive split}).

We note that under the assumption that a BGP can be answered by the union of all endpoints in $\mathcal{E}$, a trivial federation can be constructed by conjunctively splitting all triple patterns into disjunctive splits over all endpoints, i.e.
for a BGP $P=\{t_1,\ldots,t_n\}$ the \emph{trivial federation} is given by the pattern 
$\{\textrm{UNION}_{e\in\mathcal{E}}(\textrm{SERVICE}\ e \{t_1\})$, $\ldots$, $\textrm{UNION}_{e\in\mathcal{E}}(\textrm{SERVICE}\ e\{t_n\})\}$.

\textbf{Endpoint descriptions.} We note that SPARQL foresees in its standard a normative way to access a description (in the form of RDF again) from the endpoint. While the vocabulary for endpoint description is itself not part of the SPARQL standard, the accompanying VoID (vocabulary for interlinking datasets \cite{void}) standard provides respective meta-data terms to describe the endpoint's RDF dataset in terms of a natural language description, its schema (all classes and properties), as well as base statistics (overall counts of triples, classes and properties). We herein assume that endpoints provide a VoID description and/or such descriptions are automatically created through the respective tool provided as part of SPARQL-MCP.


\textbf{MCP} allows MCP-compatible LLM client LLMs (client) to initiate retrieval via the JSON-RPC interactions with a (pre-configured) MCP server. First, the client performs \textit{capability discovery} by calling the standardized \texttt{mcp\_discover} method on the server. This returns a ``manifest'' of available resources (such as documents and database records, retrieved through \texttt{get\_resource} or tool-specific equivalent passing the resource identifier, and streamed over HTTP or SSE), prompts (for interacting with language models; allowing the client to receive the fully instantiated prompt based on the predefined template), and tools (executed via the \texttt{run\_tool} or eq., passing the tool identifier and JSON-encoded arguments, and receiving the pre-defined output schema). Underpinning all retrieval operations is a stateful session preserving context (authentication tokens, intermediate results). 




\section{SPARQL-MCP}\label{sec:sparql-mcp}

In order to integrate SPARQL federation with agentic AI systems, we present SPARQL-MCP,\footnote{See \texttt{https://github.com/semantisch/sparqlmcp} for the SPARQL-MCP documentation and code (MIT License)} an MCP extension for SPARQL 1.1 federation engines that, together with schema exploration, enables AI agents to explore and query federated SPARQL endpoints. Unlike existing SPARQL-MCP implementations (see \Cref{sec:discussion} below), SPARQL-MCP specifically addresses federated querying while tackling also uneven SPARQL feature support and metadata heterogeneity (by dynamic exploration/retrieval of VoID descriptions). 

 

One notable limitation of SPARQL 1.1 includes the tendency of broadly adopted endpoints such as Wikidata and DBPedia to block SERVICE calls, with the major reason for such policy being that unconstrained SERVICE calls are prone to generating extensive result sets \cite{wu2015dynamic} and thus unacceptable processing load on the endpoint. We solve this problem similarly to \cite{moos2025sparql} by redirecting all multi-agent SERVICE queries to a proxy SPARQL federation engine, which, in turn, choreographs requests to the remote endpoints. For the purposes of modularity, concrete federation engines should be easily interchangeable, delimiting the query plan generation from the agentic endpoint discovery and exploration features. \Cref{fig:architecture} illustrates our architecture together with the protocols used for communication/data exchange between the components. We specify a particular lightweight federation engine used for evaluation in \Cref{sec:evaluation}.

\textbf{Query Execution.} For federated query execution, the SPARQL-MCP server implements the \texttt{run\_sparql\_query} tool accepting a SPARQL query string, requested result format and execution timeout. Queries must include SERVICE operators naming the target SPARQL endpoint(s). If the query contains more than one SERVICE, it is executed via a local SPARQL federation endpoint, which decomposes and forwards subqueries to respective remote endpoints. 
If the query contains exactly one SERVICE operator, the SERVICE wrapper will be removed and the query body sent directly to the endpoint (to address the blocking behaviour discussed above). 





\textbf{VoiD Computation.} Computing VoID descriptions can be resource-intensive, especially for live endpoints. Therefore, the VoiD retrieval tool (\texttt{get\_void\_descriptions}) follows a step-wise approach: (1) check cached versions of VoiD descriptions in the internal catalogue, (2) if not available, retrieve the VoID descriptions (via ``well-known'' URLs, VoID triples in the default graph, VoID in named graphs, and service-description links) from the endpoint and cache them, (3) if VoiD descriptions are not directly available, recreate parts of the VoiD descriptions following the SPORTAL approach~\cite{hasnain2016sportal} for computing VoID-style descriptions using SPARQL~1.1 self-descriptive queries, and then cache them.  



\section{FKGQA Benchmark}\label{sec:fkgqa-benchmark}

Several benchmarks exist for federated SPARQL query processing and KGQA over single knowledge graphs, but no general-purpose benchmark evaluates federated KGQA end-to-end. We extend Spider4SPARQL~\cite{kosten2023spider4sparql} (19 RDF datasets, 1,034 natural language questions, 542 SPARQL queries are publicly available) by defining a principled approach for dataset federation, cross-partitioning each dataset across multiple endpoints. Unlike existing benchmarks that assume explicit endpoint references, our benchmark requires agents to select relevant endpoints themselves, as is expected in real-world use cases.


\subsection{Partitioning Method} 


Our partitioning goal is to maximize federation patterns (conjunctive and disjunctive splits) while achieving broad coverage, balanced shard sizes, and semantic coherence. Unlike approaches that minimize cross-partition queries~\cite{azza-etal-2024SWJ}, we construct minimal partition sets such that every BGP would require federation, using three strategies: vertical (predicate-based), class-based, and horizontal (subject-hash-based) partitioning.

\textbf{Class Sharding.}
Under the assumption (which holds for all Spider4SPARQL queries when extended by axioms included in the benchmark) that all triple subjects are typed (i.e., each node appears in a $(s_d,\texttt{rdf:type},C)$ triple in the dataset, we can assign each class IRI to a shard, and place all RDF triples in $G$ whose \emph{subject} is an instance of that class into the corresponding shard. 
Formally, if a BGP mentions subject with two different types, e.g. explicitly $(s_1,\texttt{rdf:type},c_1), (s_2 \texttt{rdf:type},c_2)$, or where resp. types $c_1$ or $c_2$ are inferrable by domain or range in the dataset (e.g. $s_1, p, o_1$ in the query with \texttt{$p$,\texttt{rdfs:domain},$c_1$} in the dataset), and where $c_1 \not= c_2$ and $c_1,c_2\in I$, we split the dataset by placing all triples with subjects of these two classes in the data in different shards. 

\textbf{Predicate (Vertical) Sharding} places triples in separate shards based on their predicates. 
Formally, for a BGP mentioning triple patterns $(s_1,p_1,o_1)$, $(s_2,p_2,o_2)$ where $p_1 \not= p_2$ with $p1,p2 \in I$, we split the dataset by placing all triples with these two predicates in different shards.

\medskip Both the above methods are intended to require \emph{conjunctive splits} in the federation.

\textbf{Horizontal Sharding,} lastly, is intended to require a \emph{disjunctive split}:
here, again we exploit that all subjects are typed in the dataset, but we partition the dataset by instances (subjects) of a single class mentioned in the query: for each inferred class $c$, all subjects $s_d$ that are instances of $c$ are assigned to shards by a deterministic rule (e.g., hash of the IRI, or a skolemized blank node in place of the IRI). Every data triple $(s_d,p_d,o_d)$ is placed in the shard of $s_d$. This separates entities within a single class across shards, and is particularly suitable for queries with very small BGPs (e.g., a single triple pattern).

\textbf{Applicability Criteria.}
Let $T(q)$ be the BGP of query $q$. Define the predicate set
$\mathcal{P}(q)=\{p_q \mid (s_q,p_q,o_q)\in T§(q),\, p_q \text{ is an IRI}\}$.
Let $\mathcal{C}(q)$ be the set of class IRIs inferable from $T(q)$ via explicit \texttt{rdf:type} patterns or OWL \texttt{rdfs:domain}/\texttt{rdfs:range}. Let $\mathcal{S}(q)=\{s_q \mid (s_q,p_q,o_q)\in T(q),\, s_q \text{ is a variable}\}$ be the set of subject variables.
Then predicate  sharding is applicable iff $|\mathcal{P}(q)| \ge 2$, class sharding is applicable iff $|\mathcal{C}(q)| \ge 2$, and horizontal sharding is applicable iff $|\mathcal{S}(q)| \ge 1$ and $|\mathcal{C}(q)| \ge 1$. On the query level, across the benchmark, semantic/class sharding applies to 40.14\% of queries, vertical/predicate sharding to 95.94\%, and horizontal sharding applies to all of queries. Thus, the strategies are broadly applicable and can be combined in various ways to create diversely federated datasets. 

\textbf{Rule Selection as Set Cover.} In our implementation, we define picking sharding candidates in a manner that distributes the dataset graph $G$ over a minimal set of shards, selecting applicable candidate shardings for each query, which we can formalize as a set-cover problem (NP-hard) \cite{karp2009reducibility} per dataset. As $Q$ be the set of queries, let $\mathcal{R}$ be the set of applicable shardings (predicate, class and horizontal splits). Each $r\in R$ then covers a number of queries in $Q$. The objective is to select a minimum-size subset $\mathcal{R}^\ast \subseteq \mathcal{R}$ such that $\bigcup_{r \in \mathcal{R}^\ast} C(r) = Q$. Formally we pick:
\[
\min\; |\mathcal{R}^\ast|\quad \text{s.t.}\quad \forall q \in Q,\; \exists r \in \mathcal{R}^\ast: q \in C(r).
\]

Since class shardings are naturally dominated by horizontal, we omit horizontal shardings as candidates for any query where a class sharding is applicable, ensuring higher variety in the federation patterns. We can then encode this set-cover formulation as an Answer Set Programming (ASP) problem~\cite{brewka_etal_asp} to compute such a minimal set of shardings per dataset.\footnote{See \url{https://github.com/semantisch/fkgqa} for the benchmark dataset, documentation and all code for generating the benchmark (MIT License)}

\textbf{Shard Materialisation.}
Given a selected sharding set, we materialize shards by assigning each data triple $(s_d,p_d,o_d)\in G$ to the shard dictated by the sharding approaches (predicate-based for vertical shardings, class-of-$s_d$ for class shardings, and class-conditioned hash of $s_d$ for horizontal shardings), and place all remaining triples in a base shard.

 
\subsection{Resulting Benchmark}

We materialise federated shards for 19 datasets, yielding 118 shards in total, each of which are hosted in a separate endpoint. The minimal sharding-set size averages 3.7 (median 4), ranging from 2 to 8 rules per dataset, and the sharding composition becomes 18.30\% vertical, 30.42\% semantic, and 51.28\% horizontal, which, we hypothesise, is an inverse of the ranking of partitioning approaches to \emph{minimize} query costs in proposed federated systems (see \Cref{sec:related-work}). The number of shards per dataset ranges from 3 to 14 (median 6, mean 6.21). Shard balance (as the average coefficient of variation of shard sizes per dataset) is 0.88, with a moderate skew while keeping shards in comparable magnitudes of triples.

\textbf{Federation Coverage.}
By construction, the selected shardings federate every query. We define the \emph{fan-out} $f(Q)$ as the number of distinct shards/endpoints a BGP query $Q=\{t_1,\ldots,t_n\}$ may need to contact in order to ensure complete results across all endpoints in $\mathcal{E}$ as follows: let $\mathcal{E}_t$ be the set of all endpoints in $\mathcal{E}$ returning a non-empty result on triple pattern $t$, then we define $f(Q)=|\bigcup_{t\in Q}\mathcal{E}_t|$. Note that $f(Q)$ is an upper bound, i.e. based on the joins and underlying data, $Q$ may be correctly/completely answered using less endpoints as shown by \Cref{ex:2}.

\begin{example}\label{ex:2} Example BGP for the question ``How many ships ended up being `Captured'?'':\vspace{2px} \\
$(?s,\texttt{rdf:type},\texttt{:ship})$ and $(?s,\texttt{:ship\#disposition\_of\_ship},\texttt{"Captured"})$.

\end{example}

Under horizontal sharding on \texttt{:ship}, ship instances are split across multiple shards. However, in the materialized data all \texttt{"Captured"} ships hash to \texttt{shard \#5}, so the query touches only that shard. In the evaluation (see \Cref{sec:evaluation}), we additionally explore whether all potentially matching shards are actually consulted by the agent.
    
Based on shardings-implied fan-out, queries touch 6.48 shards on average (median 6, min 2, max 14). As shown by the example, in practice, realized fan-out can be lower: 4.84\% have no matches and 24.49\% queries match exactly one shard in realisation.



\textbf{VoiD Computation.}
We pre-compute VoID descriptions for all 118 shards using the SPARQL-MCP \texttt{get\_void\_descriptions} tool, providing both standard and extended (with linksets) variants.

\section{Evaluation}\label{sec:evaluation}



We implement the evaluation suite using the Agents SDK.\footnote{See \url{https://github.com/openai/openai-agents-python}} Tasks are executed by a ReAct-style agent that alternates \emph{LLM calls} and \emph{tool calls}. We structure the evaluation around an ablation design, looking into performance in the baseline case (endpoint URLs provided, federated querying tool available), basic high-level descriptions scenario and the full scenario with both federated querying and void retrieval tools available. For all evaluation setups, we limit turns to 10 per evaluation run, and log every step. We evaluate \texttt{GPT-5.2} (\texttt{gpt\_5\_2\_2025\_12\_11}) as a high-capability frontier model and \texttt{Qwen3-8B} (\texttt{qwen\_8b}) as a reproducible open-weight baseline representative of compact agent-capable models~\cite{yang2025qwen3}. As hyperparameters, we set \texttt{temperature} and thinking-related hyperparameters to the lowest values.

Runs yielding failed, unsound, or incomplete results are counted as evaluation errors and final query results are matched to the ground truth results based on signature sets. Prior to each evaluation, we issue a ramp\,\,up mix to warm caches and exclude cold-start effects. All experiments ran on Debian 13 server with 16 CPUs (Intel Xeon), 128\,GB RAM; Python\,3.13 and GCC\,14.2.\footnote{See the full evaluation results, evaluation scripts and the respective documentation (MIT License): \url{https://github.com/semantisch/sparql-mcp-evaluation}}

\textbf{Evaluation Setups.} We evaluate in three levels: (i) \texttt{baseline}: the agent only knows the endpoint URLs and has access to the \texttt{run\_sparql\_query} tool; (ii) \texttt{high\_level}: the agent receives URLs plus one-sentence, topical natural-language endpoint descriptions,\footnote{E.g., "A catalog of automobile models associated with their manufacturers." and "A reference-style catalog of automobile makes and models with associated geographic origin information." All high-level descriptions are included in the benchmark repository.} and has access to the \texttt{run\_sparql\_query} tool; (iii) \texttt{void\_tool}: \texttt{baseline}, plus the agent has access to the \texttt{get\_void\_descriptions} tool for schema exploration. In all conditions, agents must discover endpoints, synthesize federated SPARQL queries, execute via \texttt{run\_sparql\_query}, and refine based on results, submitting the final query via the evaluation environment-introduced \texttt{submit\_final\_query} call.

\textbf{Evaluation Metrics.}
As results addressing various aspects of the agentic SPARQL performance, we report (i) a descriptive analysis of execution traces, (ii) syntactic validity of generated SPARQL queries, (iii) overall pipeline accuracy, (iv) an evaluation of endpoint accuracy (i.e., assessing which endpoints are consulted by the agent), and, finally, (iv) assess distinctive behavioural patterns.

\section{Results}\label{sec:results}

 \textbf{Descriptive Analysis.} In total, the evaluation has included 5,814 separate agent runs. In terms of cost, a total of 82,574,734 tokens has been used. Of these, GPT-5.2 contributed 61,429,831 tokens (median 20,480 tokens per run, min.\ 2,457, max.\ 172,614), while Qwen3-8B contributed 21,144,903 tokens in total (median 8,288 tokens per run, min.\ 1,550, max.\ 335,299), illustrating that the higher-capacity model dominates overall token usage. In total, end-to-end runtime of all agent runs has been above 65 hours, with a median runtime of 22.2 seconds per run (max.\ 11 minutes). Breaking this down by model, GPT‑5.2 achieved a median end‑to‑end runtime of 16.3 s. per run, whereas Qwen3‑8B required almost twice as much, 31.9 s. In terms of call patterns, GPT‑5.2 averaged 5.5 LLM calls and 9.0 SPARQL calls per run, whereas Qwen3‑8B averaged 7.2 LLM calls and 6.4 SPARQL calls per run; VoID retrieval calls averaged 1.0–1.1 per run in the respective setup, with SPARQL call counts highest in the baseline (11.1 for GPT‑5.2, 6.4 for Qwen3‑8B) and reduced to 6.2 for GPT‑5.2 when high‑level endpoint descriptions were provided, while Qwen3‑8B showed similar counts (6.8) in that mode.

\textbf{Syntactic Analysis.} Syntactic analysis shows that 29,431 out of 38,886 SPARQL queries (75.7\% overall, across all LLMs and setups) executed successfully, i.e., were syntactically correct. GPT‑5.2 achieved high success rates (97.4\%--98.0\% across setups), while Qwen3‑8B showed lower rates (41.5\%--61.1\%), with the highest rate in the \texttt{high\_level} setup. For Qwen3‑8B, the primary pitfalls were: 41.6\% failed parsing, 24.3\% had prefix issues, 1.3\% had unmatched braces, and 1.5\% had malformed SERVICE syntax.

\textbf{Final Results Analysis.} GPT‑5.2 achieved accuracy rates of 42.1\% in \texttt{baseline}, 45.4\% in \texttt{high\_level}, and 43.5\% in \texttt{void\_tool}. By contrast, Qwen3‑8B achieved accuracy rates of 13.1\% in \texttt{baseline}, 13.2\% in \texttt{high\_level}, and 13.8\% in \texttt{void\_tool}. For the larger LLM, we could, therefore, in the federated setup, achieve the exact same success rates as the state-of-the-art (including LLM-driven) approaches evaluated on the original benchmark (45\%) \cite{kosten2023spider4sparql}.

\textbf{Endpoint Accuracy.} GPT‑5.2 achieved high consultation rates in \texttt{baseline} (90.7\% for successful) and \texttt{void\_tool} (91.7\% for successful), but lower rates in \texttt{high\_level} (25.8\% for successful), suggesting that high-level descriptions enable effective source pre-selection. Qwen3‑8B showed consistently high rates in \texttt{void\_tool} (98.6\% for successful). A large fraction of queries were \emph{trivial federations} (consulting all endpoints when fewer would suffice): GPT‑5.2 produced trivial queries in 90.2\%--91.7\% of successful runs in baseline modes but only 11.0\% in \texttt{high\_level}, while Qwen3‑8B showed high trivial query rates across all modes (68.5\%--98.6\%), indicating a tendency to query all endpoints rather than performing selective discovery.

\textbf{Behavioural Analysis.} 
Behavioral analysis reveals distinct agent strategies: GPT‑5.2 exhibits exploration behavior with frequent endpoint switching and minimal query redundancy, gradually discovering endpoints. By contrast, Qwen3‑8B shows almost no exploration, rarely switches endpoints, but exhibits high query redundancy, discovering endpoints quickly by querying all endpoints simultaneously. GPT‑5.2 achieves faster time-to-first-result compared to Qwen3‑8B, while both models show similar correction dynamics.

\section{Discussion}\label{sec:discussion}

Our evaluation demonstrates that agentic federated SPARQL querying is viable for sufficiently capable models: GPT-5.2 achieved accuracy rates (42.1\%--45.4\%) comparable to state-of-the-art approaches on the original Spider4SPARQL benchmark (45\%)~\cite{kosten2023spider4sparql}, despite the added complexity of federation. High-level endpoint descriptions proved more effective than detailed VoID metadata for guiding source selection, reducing trivial queries from 90.2\% to 11.0\% for GPT-5.2 while improving accuracy. However, smaller models (Qwen3-8B) showed significantly lower performance (13.1\%--13.8\% accuracy) and high syntactic error rates (41.5\%--61.1\%), indicating that model scale and SPARQL-specific capabilities are still critical. The behavioral differences such as GPT-5.2's exploration strategy versus Qwen3-8B's approach of querying all endpoints simultaneously suggest that different models require tailored tool designs and prompting guidance. While LLMs can leverage semantic understanding for endpoint selection, they often lack the cost-awareness of traditional federated query optimizers, leading to inefficient query plans that consult unnecessary endpoints (trivial plans).

\section{Related Work}\label{sec:related}\label{sec:related-work}

\textbf{Federated Query Processing}\label{subsec:federated-sparql} enables data integration across distributed endpoints. Since federated execution must account for heterogeneous source capabilities and incomplete or imprecise metadata (and, to a lesser extent, network effects), substantial work has focused on query optimisation. FedX~\cite{fedx} introduced query decomposition and source selection techniques, while CoDA~\cite{Coda} proposed cost-based strategies for aggregates and alternative join plans. Heuristic reordering approaches~\cite{Yannakis2018HeuristicsbasedQR} further improve performance without requiring complete statistics, and Heling and Acosta~\cite{heling2022federated} extended optimisation to heterogeneous Linked Data Fragment (LDF) interfaces via interface-aware physical operators.

Endpoint discovery and schema exploration are central to federated querying, with prior work covering metadata-driven, ad-hoc query-based, and hybrid discovery strategies~\cite{montoya2017odyssey,heling2019quality}, as well as VoID-based and probing-based schema inference~\cite{zeimetz2019analyzing,heling2022federated}. Evaluation of federated SPARQL querying rely on benchmarks including FedBench~\cite{schmidt2011fedbench}, QFed~\cite{rakhmawati2014qfed}, and LargeRDFBench~\cite{saleem2018largerdfbench}. Nonetheless, most federated SPARQL systems require manual query formulation, limiting accessibility. Recent efforts have aimed to bridge the gap between natural language interfaces and federated SPARQL querying, with Emonet et al.~\cite{emonet2024} introducing a RAG framework that translates natural language questions into federated SPARQL queries over bioinformatics knowledge graphs. Widely used single-source KGQA benchmarks include QALD~\cite{unger2012template}, LC-QuAD~\cite{trivedi2017lc}, and Spider4SPARQL~\cite{kosten2023spider4sparql}.

\textbf{KG Schema Exploration}\label{subsec:rag-schema-llm}
Prior work in KGQA highlights structure- and schema-aware exploration, including agentic approaches~\cite{gu-etal-2023-dont}, iterative reasoning methods~\cite{sun2023think,luo2023reasoning,chen2024plan}, ranking and pruning strategies~\cite{tian2024augmenting,tan2025paths}, graph-based importance measures~\cite{jimenez2024hipporag,gutierrez2025rag}, AMR-guided mappings~\cite{emonet2024}, and learned retrievers~\cite{li2024simple}. Complementary work pre-computes indexes for schema access~\cite{walter2025graspgenericreasoningsparql}. While effective for single-KG settings, these methods assume a fixed, known graph. Our work focuses on endpoint- and schema-level exploration in federated settings, exposing metadata (e.g., VoID descriptions) operationally via MCP.

\textbf{MCP and Federated Retrieval.} MCP wrappers have been proposed for SPARQL. The Wikidata SPARQL MCP Server\footnote{See \url{https://github.com/QuentinCody/wikidata-sparql-mcp-server}} targets the Wikidata endpoint and exposes a single \texttt{sparql\_query} tool. Triplestore vendors also provide MCP servers, including the Apache Jena MCP Server\footnote{See \url{https://github.com/ramuzes/mcp-jena}}, compatible with Jena Fuseki\footnote{See \url{https://jena.apache.org/documentation/fuseki2/fuseki-docker.html}}, as well as corresponding solutions for GraphDB\footnote{See \url{https://graphdb.ontotext.com/documentation/11.1/using-graphdb-llm-tools-with-external-clients.html}} and Stardog\footnote{See \url{https://lobehub.com/mcp/vairpower-stardog_graphrag_mcp_poc}}; these primarily support direct query/update operations and basic graph inspection. The RDF Explorer MCP server\footnote{See \url{https://github.com/emekaokoye/mcp-rdf-explorer}} additionally offers endpoint-specific statistics and search utilities. To the best of our knowledge, none of these implementations support schema exploration or federated SPARQL querying, nor has prior work systematically evaluated ReAct-style MCP-based SPARQL agents.


\section{Conclusions and Future Work}\label{sec:conclusions}

This work introduces agentic federated SPARQL querying, enabling LLM-based agents to autonomously discover endpoints, explore schemas, and formulate federated queries from natural language questions. Our contributions include: (i) a combined FKGQA benchmark for evaluating agentic federated Knowledge Graph Question Answering; (ii) the SPARQL-MCP server implementation with tools for federated querying, endpoint discovery, and schema exploration; and (iii) a comprehensive evaluation across 5,814 agent runs comparing different architectural options. Our evaluation reveals that agentic federated SPARQL querying is viable for sufficiently capable models: GPT-5.2 achieved accuracy rates (42.1\%--45.4\%) comparable to state-of-the-art approaches on the original Spider4SPARQL benchmark (45\%)~\cite{kosten2023spider4sparql}, despite the added complexity of federation. High-level endpoint descriptions proved more effective than detailed VoID metadata, reducing trivial queries from 90.2\% to 11.0\% while improving accuracy. However, smaller models showed significantly lower performance, indicating that model scale and SPARQL-specific capabilities are critical for production use. Below, we outline important directions for future work:

\begin{description}
    \item[FW1: SPARQL Capabilities:] Addressing \textbf{C5} discussed in \Cref{sec:intro}, the significantly lower syntactic validity rates for Qwen3-8B (41.5\%--61.1\% vs.\ 97.4\%--98.0\% for GPT-5.2) indicate that smaller models require substantial improvements in SPARQL-specific fine-tuning or more structured query generation pipelines to be viable.
    
    \item[FW2: Schema Exploration:] Even beyond inherent metadata heterogeneity (\textbf{C3}), enabling the VoID tool did not consistently improve accuracy over high‑level descriptions, consultation rates in \texttt{void\_tool} (91.7\% for GPT-5.2, 98.6\% for Qwen3-8B) or trivial-query rates. This finding contrasts with VoID-based schema inference work~\cite{zeimetz2019analyzing,heling2022federated}, which assumed more sophisticated planning. In agentic settings, simpler high-level descriptions may be more effective, which prompts future work to test how to structure metadata for agents.
    
    \item[FW3: SPARQL Registry:] As thematised in \textbf{C4} and \textbf{C2}, real-world endpoints face unavailability, latency and timeouts, have heterogeneous interfaces, and missing metadata while catalogues (e.g., SPARQLES \cite{vandenbussche2017sparqles}) are largely inaccessible. Future work should extend catalogue metadata (partitions, availability/quality) and provide searchable, up-to-date endpoint and ontology repositories.
    
\end{description}

\subsubsection*{Acknowledgements}This research was funded in whole or in part by the Austrian Science Fund (FWF) Cluster of Excellence ``Bilateral AI'' (BILAI) [10.55776/COE12].

\pagebreak
\bibliographystyle{named}

\end{document}